# The three-dimensional Dirac-Oscillator in the presence of Aharonov-Bohm and magnetic monopole potentials

A. D. Alhaidari

*Physics Department, King Fahd University of Petroleum & Minerals, Dhahran 31261, Saudi Arabia*
e-mail: haidari@mailaps.org

We study the Dirac equation in 3+1 dimensions with non-minimal coupling to isotropic radial three-vector potential and in the presence of static electromagnetic potential. The space component of the electromagnetic potential has angular (non-central) dependence such that the Dirac equation is completely separable in spherical coordinates. We obtain exact solutions for the case where the three-vector potential is linear in the radial coordinate (Dirac-Oscillator) and the time component of the electromagnetic potential vanishes. The relativistic energy spectrum and spinor eigenfunctions are obtained.

## I. INTRODUCTION

The objective of adding a three-vector potential, which is linear in the coordinate, to the Dirac equation in an analogy to the kinetic energy term which is linear in the momentum lead Moshinsky and Szczepaniak to the solution of the Dirac-Oscillator problem [1]. The nonrelativistic limit reproduces the usual Harmonic oscillator. The linear potential had to be added, via non-minimal coupling, to the odd part of the Dirac operator. Subsequently, the Dirac-Oscillator attracted a lot of attention [2]. Our contribution here is to study the same relativistic problem but in the presence of Aharonov-Bohm and magnetic monopole potentials. This problem does not have the simplicity of spherical symmetry where the solution of the radial Dirac equation is the only concern. The angular dependence of the Dirac operator in this problem is highly non-trivial and one may attempt to separate the angular variables into an "angular Dirac equation" and search for its solutions. Ferkous and Bounames solved the Dirac-Oscillator problem in two dimensions in the presence of Aharonov-Bohm effect [3]. Villalba and Maggiolo obtained the energy spectrum of the two dimensional Dirac-Oscillator in the presence of a constant magnetic field [4]. However, no report on the solution of the problem in three dimensions was presented. In fact, little work has been done on the Dirac equation in three dimensions where the electromagnetic potential has angular variations and the solution requires separation of variables [5]. Moreover, most of the work on separation of variables in the Dirac equation concentrated on separation in curved space-times with gravitational coupling and in the absence of the electromagnetic potential with the exception of very few. For the latest publications in this topic with citation to earlier work, one may consult the papers listed in [6]. Recently, we have introduced a systematic and intuitive approach for the separation of variables in spherical coordinates for the Dirac equation with coupling to static non-central electromagnetic potential [7]. The Dirac-Coulomb problem in the presence of the Aharonov-Bohm effect and magnetic monopole potential was solved. The relativistic bound states energy spectrum and corresponding spinor wavefunctions were obtained. In the present article the same formulation will be extended to the relativistic Dirac-Oscillator problem.

The three-dimensional Dirac equation for a free structureless particle of spin $\frac{1}{2}$ reads $\left(i\hbar\gamma^{\mu}\partial_{\mu}-mc\right)\psi=0$, where $m$ is the rest mass of the particle, $c$ is the speed of light, and $\psi$ is a four-component wavefunction. The four matrices $\{\gamma^{\mu}\}_{\mu=0}^{3}$ are given the following standard representation

$$\gamma^{0}=\begin{pmatrix}1 & 0\\ 0 & -1\end{pmatrix},\ \vec{\gamma}=\begin{pmatrix}0 & \vec{\sigma}\\ -\vec{\sigma} & 0\end{pmatrix}, \tag{1.1}$$

where $1$ is the 2×2 unit matrix and $\vec{\sigma}$ are the usual 2×2 Pauli spin matrices. In the atomic units, $\hbar = m = 1$, the Dirac equation reads $\left(i\gamma^{\mu}\partial_{\mu}-\lambdabar^{-1}\right)\psi=0$, where $\lambdabar = \hbar/mc$ = $1/c$ is the Compton wavelength of the particle. In the presence of the electromagnetic potential, $A_{\mu}=(A_{0},c\vec{A})$, gauge invariant coupling to the charged spinor is accomplished by the "minimal substitution" $\partial_{\mu}\to\partial_{\mu}+i\lambdabar A_{\mu}$, which transforms the free Dirac equation into $\left[i\gamma^{\mu}(\partial_{\mu}+i\lambdabar A_{\mu})-\lambdabar^{-1}\right]\psi=0$. Written in details, it reads as follows

$$i\lambdabar\frac{\partial}{\partial t}\psi=\left(-i\vec{\alpha}\cdot\vec{\nabla}+\vec{\alpha}\cdot\vec{A}+\lambdabar A_{0}+\lambdabar^{-1}\gamma^{0}\right)\psi, \tag{1.2}$$

where $\vec{\alpha}=\gamma^{0}\vec{\gamma}=\begin{pmatrix}0 & \vec{\sigma}\\ \vec{\sigma} & 0\end{pmatrix}$. Now, we introduce coupling to the spherically symmetric radial three-vector potential $\vec{B}(\vec{r})=W(r)\hat{r}$ by the non-minimal substitution $\vec{\nabla}\to\vec{\nabla}+\vec{B}\gamma^{0}$ (i.e., $\vec{P}\to\vec{P}-iW(r)\hat{r}\gamma^{0}$, where $\vec{P}$ is the linear three-momentum operator $-i\vec{\nabla}$). Thus, we obtain

$$i\frac{\partial}{\partial t}\psi=\left[-i\lambdabar\vec{\alpha}\cdot\left(\vec{\nabla}+W\hat{r}\gamma^{0}\right)+\lambdabar\vec{\alpha}\cdot\vec{A}+\lambdabar^{2}A_{0}+\gamma^{0}\right]\psi, \tag{1.3}$$

where $\hat{r}$ is the radial unit vector and $W(r)$ is a real radial function. For time independent potentials, equation (1.3) gives the following matrix representation of the Dirac Hamiltonian (in units of $mc^{2}=\lambdabar^{-2}$)

$$\mathcal{H}=\begin{pmatrix}1+\lambdabar^{2}A_{0} & -i\lambdabar\vec{\sigma}\cdot\vec{\nabla}+\lambdabar\vec{\sigma}\cdot\vec{A}+i\lambdabar W\vec{\sigma}\cdot\hat{r}\\ -i\lambdabar\vec{\sigma}\cdot\vec{\nabla}+\lambdabar\vec{\sigma}\cdot\vec{A}-i\lambdabar W\vec{\sigma}\cdot\hat{r} & -1+\lambdabar^{2}A_{0}\end{pmatrix}. \tag{1.4}$$

Thus the energy eigenvalue equation reads $(\mathcal{H}-\varepsilon)\psi=0$, where $\varepsilon$ is the relativistic energy which is real and measured in units of $mc^{2}$.

In Sec. II, we follow the approach developed in [7] for the separation of variables of the three dimensional Dirac equation in spherical coordinates. The solution space for the angular component of the Dirac equation is constructed for the case where the space component of the electromagnetic potential is a combination of Aharonov-Bohm and magnetic monopole potentials. In Sec. III, we obtain the solution of the radial Dirac equation for the case where the time component of the electromagnetic potential vanishes and $W(r)$ is linear in $r$. The relativistic energy spectrum will be obtained in the same section. In Sec. IV, we show that the solution of the spherically symmetric problem is a special case of our findings. Moreover, we obtain the nonrelativistic limit and verify that it reproduces earlier results.

## II. SOLUTION OF THE ANGULAR EQUATION

We take the time-independent electromagnetic potential to have the following components in spherical coordinates

$$A_0(\vec{r}) = V(r), \quad \vec{A}(\vec{r}) = \hat{\phi}[U(\theta)/r\sin\theta], \tag{2.1}$$

where $U$ and $V$ are real functions. This potential satisfies the Lorentz gauge fixing condition $\partial_\mu A^\mu = 0$ as well as the radiation gauge, $\vec{\nabla}\cdot\vec{A} = 0$. If we write the spinor wave function as $\psi(\vec{r}) = \frac{1}{r\sqrt{\sin\theta}}\begin{pmatrix} if_+(\vec{r}) \\ f_-(\vec{r}) \end{pmatrix}$, then the action of the Dirac Hamiltonian (1.4) on the four-component spinor $\begin{pmatrix} f_+ \\ f_- \end{pmatrix}$ is represented by the 4×4 matrix operator

$$\mathcal{H} = 1 \otimes \mathcal{H}_0 + \lambdabar\vec{\sigma}\cdot\hat{r}\otimes\mathcal{H}_r + \lambdabar\frac{\vec{\sigma}\cdot\hat{\theta}}{r}\otimes\mathcal{H}_\theta + \lambdabar\frac{\vec{\sigma}\cdot\hat{\phi}}{r\sin\theta}\otimes\mathcal{H}_\phi, \tag{2.2}$$

where $(\hat{r},\hat{\theta},\hat{\phi})$ are the unit vectors in spherical coordinates and

$$\mathcal{H}_0 = \begin{pmatrix} 1+\lambdabar^2 V(r) & 0 \\ 0 & -1+\lambdabar^2 V(r) \end{pmatrix}, \quad \mathcal{H}_r = \begin{pmatrix} 0 & -\partial_r + \frac{1}{r} + W(r) \\ \partial_r - \frac{1}{r} + W(r) & 0 \end{pmatrix}, \tag{2.3a}$$

$$\mathcal{H}_\theta = \begin{pmatrix} 0 & -\partial_\theta + \frac{\cos\theta}{2\sin\theta} \\ \partial_\theta - \frac{\cos\theta}{2\sin\theta} & 0 \end{pmatrix}, \quad \mathcal{H}_\phi = \begin{pmatrix} 0 & -\partial_\phi - iU(\theta) \\ \partial_\phi + iU(\theta) & 0 \end{pmatrix}. \tag{2.3b}$$

Square integrability and the boundary conditions require that the components of the spinor wavefunction satisfy $f_\pm(r)\big|_{\substack{r=0 \\ r\to\infty}} = 0$, $f_\pm(\theta)\big|_{\substack{\theta=0 \\ \theta=\pi}}$ is finite, and $f_\pm(\phi) = f_\pm(\phi+2\pi)$.

To simplify the solution of the problem, we search for a local 2×2 unitary transformation matrix $\Lambda(\vec{r})$ that maps the spherical projection of the Pauli matrices $(\vec{\sigma}\cdot\hat{\theta}, \vec{\sigma}\cdot\hat{\phi}, \vec{\sigma}\cdot\hat{r})$ into their canonical Cartesian representation $(\sigma_1, \sigma_2, \sigma_3)$, respectively. That is,

$$\Lambda^{-1}\vec{\sigma}\cdot\hat{\theta}\Lambda = \sigma_1, \quad \Lambda^{-1}\vec{\sigma}\cdot\hat{\phi}\Lambda = \sigma_2, \quad \Lambda^{-1}\vec{\sigma}\cdot\hat{r}\Lambda = \sigma_3. \tag{2.4}$$

Other permutations of the $\sigma_i$'s on the right are equivalent, differing only by a unitary transformation. The following unitary matrix fulfills (2.4)

$$\Lambda(\vec{r}) = e^{-\frac{i}{2}\sigma_3\phi} e^{-\frac{i}{2}\sigma_2\theta}. \tag{2.5}$$

The transformed wavefunction, which we write as $\chi = \begin{pmatrix} g_+ \\ g_- \end{pmatrix}$, has the two-component spinors $g_\pm = \Lambda^{-1} f_\pm$, whereas the matrix operator (2.2) gets mapped into the following hermitian representation of the Dirac Hamiltonian

$$H = 1 \otimes H_0 + \lambdabar\sigma_3 \otimes H_r + \lambdabar\frac{\sigma_1}{r}\otimes H_\theta + \lambdabar\frac{\sigma_2}{r\sin\theta}\otimes H_\phi, \tag{2.6}$$

where,

$$H_0 = \mathcal{H}_0, \quad H_r = \begin{pmatrix} 0 & -\partial_r + W(r) \\ \partial_r + W(r) & 0 \end{pmatrix}, \tag{2.7a}$$

$$H_\theta = \begin{pmatrix} 0 & -\partial_\theta + \sigma_3\frac{U(\theta)}{\sin\theta} \\ \partial_\theta - \sigma_3\frac{U(\theta)}{\sin\theta} & 0 \end{pmatrix}, \quad H_\phi = \begin{pmatrix} 0 & -\partial_\phi \\ \partial_\phi & 0 \end{pmatrix}, \tag{2.7b}$$

and we have used,

$$\Lambda^{-1}\partial_r\Lambda = \partial_r, \quad \Lambda^{-1}\partial_\theta\Lambda = \partial_\theta - \frac{i}{2}\sigma_2, \quad \Lambda^{-1}\partial_\phi\Lambda = \partial_\phi + \frac{i}{2}(\sigma_1\sin\theta - \sigma_3\cos\theta). \tag{2.8}$$

Finally, we obtain the following complete Dirac equation $(H-\varepsilon)\chi = 0$:

$$\left[\begin{pmatrix} 1+\lambdabar^2 V-\varepsilon & \lambdabar\sigma_3(-\partial_r+W) \\ \lambdabar\sigma_3(\partial_r+W) & -1+\lambdabar^2 V-\varepsilon \end{pmatrix} + \frac{\lambdabar}{r}\begin{pmatrix} 0 & -\sigma_1\partial_\theta-i\sigma_2\frac{U}{\sin\theta} \\ \sigma_1\partial_\theta+i\sigma_2\frac{U}{\sin\theta} & 0 \end{pmatrix}\right.$$
$$\left.+\frac{\lambdabar}{r\sin\theta}\begin{pmatrix} 0 & -\sigma_2\partial_\phi \\ \sigma_2\partial_\phi & 0 \end{pmatrix}\right]\begin{pmatrix} g_+ \\ g_- \end{pmatrix} = 0 \tag{2.9}$$

where $g_\pm = \begin{pmatrix} g_\pm^+ \\ g_\pm^- \end{pmatrix}$. If we write these spinor components as $g_s^\pm(\vec{r}) = R_s^\pm(r)\Theta_s^\pm(\theta)\Phi_s^\pm(\phi)$, where $s$ is the $+$ or $-$ sign, then Eq (2.9) gets completely separated in all three coordinates as follows

$$\pm\sigma_2 \frac{d}{d\phi}\Phi_\pm = \pm i\sigma_2\varepsilon_\phi\Phi_\pm, \tag{2.10a}$$

$$\left(\pm\sigma_1 \frac{d}{d\theta} \pm i\sigma_2 \frac{U+\varepsilon_\phi}{\sin\theta}\right)\Theta_\pm = \sigma_3\varepsilon_\theta\Theta_\pm, \tag{2.10b}$$

$$\begin{pmatrix} 1+\lambdabar^2 V - \varepsilon & \lambdabar\sigma_3\left(-\frac{d}{dr}+\frac{\varepsilon_\theta}{r}+W\right) \\ \lambdabar\sigma_3\left(\frac{d}{dr}+\frac{\varepsilon_\theta}{r}+W\right) & -1+\lambdabar^2 V-\varepsilon \end{pmatrix}\begin{pmatrix} R_+ \\ R_- \end{pmatrix} = 0, \tag{2.10c}$$

where $\varepsilon_\phi$ and $\varepsilon_\theta$ are the separation constants which are real and dimensionless. For the remainder of this section we repeat briefly (for ease of reference) the same development carried out in Sec. II of [7] to obtain the solution of the angular equations (2.10a,b).

Equation (2.10a) could be rewritten as $\frac{d}{d\phi}\Phi_\pm = i\varepsilon_\phi\Phi_\pm$. Therefore, its normalized solution is

$$\Phi_s^\pm(\phi) = \frac{1}{\sqrt{2\pi}}e^{i\varepsilon_\phi\phi}. \tag{2.11}$$

The requirement that $f_\pm(\phi) = f_\pm(\phi+2\pi)$ puts a restriction on the real values of $\varepsilon_\phi$. Now, $f_\pm(\vec{r}) = \Lambda(\vec{r})g_\pm(\vec{r})$, that is $f_\pm = \frac{e^{i\varepsilon_\phi\phi}}{\sqrt{2\pi}}e^{-\frac{i}{2}\sigma_3\phi}e^{-\frac{i}{2}\sigma_2\theta}\begin{pmatrix} R_\pm^+\Theta_\pm^+ \\ R_\pm^-\Theta_\pm^- \end{pmatrix}$. Therefore, we obtain the requirement that $e^{i2\pi\varepsilon_\phi} = -1$. Hence, we should have

$$\varepsilon_\phi = \frac{m}{2}, \qquad m = \pm 1, \pm 3, \pm 5, \ldots \tag{2.12}$$

Equation (2.10b), on the other hand, could be rewritten as $\left(\frac{d}{d\theta} - \sigma_3\frac{U+\varepsilon_\phi}{\sin\theta}\right)\Theta_\pm = \mp i\sigma_2\varepsilon_\theta\Theta_\pm$. In explicit matrix form it reads

$$\begin{pmatrix} \frac{d}{d\theta}-\frac{U+\varepsilon_\phi}{\sin\theta} & s\varepsilon_\theta \\ -s\varepsilon_\theta & \frac{d}{d\theta}+\frac{U+\varepsilon_\phi}{\sin\theta} \end{pmatrix}\begin{pmatrix} \Theta_s^+ \\ \Theta_s^- \end{pmatrix} = 0, \tag{2.13}$$

where again the sign $s = \pm$. This matrix equation decouples, for each angular spinor component, into a second order differential equation, which could be written in terms of the coordinate $x = \cos\theta$ as follows

$$\left[(1-x^2)\frac{d^2}{dx^2} - x\frac{d}{dx} \pm \frac{dU}{dx} - \frac{(U+\varepsilon_\phi)(U+\varepsilon_\phi \mp x)}{1-x^2} + \varepsilon_\theta^2\right]\Theta_s^\pm = 0. \tag{2.14}$$

The solution of this equation is obtained in terms of the Jacobi polynomial $P_n^{(\mu,\nu)}(x)$ as follows [7]

$$\Theta_s^\pm(\theta) = A_n(1-x)^\alpha(1+x)^\beta P_n^{(\mu,\nu)}(x), \tag{2.15}$$

where $n = 0,1,2,...$ and the normalization constant is

$$A_n = \sqrt{\frac{2n+\mu+\nu+1}{2^{\mu+\nu+1}} \frac{\Gamma(n+1)\Gamma(n+\mu+\nu+1)}{\Gamma(n+\mu+1)\Gamma(n+\nu+1)}} \ . \tag{2.16}$$

The real parameters $\mu, \nu > -1$ and due to the factor $1/\sqrt{\sin\theta}$ in the spinor wavefunction $\psi(\vec{r})$ then square integrability requires that $\alpha, \beta \geq \frac{1}{4}$. This is so because $1/\sqrt{\sin\theta} = (1-x)^{-\frac{1}{4}}(1+x)^{-\frac{1}{4}}$. Moreover, the differential equation of the Jacobi polynomial [8] dictates that the angular potential function $U$ be linear in $x$. That is, $U(\theta) = a - bx$, where $a$ and $b$ are dimensionless physical parameters. This gives the azimuth component of the electromagnetic potential as $A_\phi = \frac{a - b\cos\theta}{r\sin\theta}$, which is a combination of Aharonov-Bohm potential whose magnetic flux strength is $2\pi|a-b|$ and a magnetic monopole potential with strength $b$ and singularity along the negative $z$-axis. Additionally, we obtain [7]

$$\alpha = \frac{\mu}{2} + \frac{1}{4}, \ \beta = \frac{\nu}{2} + \frac{1}{4}, \tag{2.17a}$$

$$\mu = \left|a - b + \frac{m\mp 1}{2}\right|, \ \nu = \left|a + b + \frac{m\pm 1}{2}\right|, \tag{2.17b}$$

$$\varepsilon_\theta^2 = \left(n + \frac{1}{2}\left|a - b + \frac{m\mp 1}{2}\right| + \frac{1}{2}\left|a + b + \frac{m\pm 1}{2}\right| + \frac{1}{2}\right)^2 - b^2 . \tag{2.17c}$$

The top and bottom sign in these formulas goes with the corresponding superscript on the angular spinor component $\Theta_s^\pm$. Separability of the Dirac equation requires that $\varepsilon_\theta$ be the same for the two components, $\Theta_s^\pm$. This requirement is satisfied only if either one of the following two conditions is met

$$b = 0, \text{ or } \left|a + \frac{m}{2}\right| \geq |b| + \frac{1}{2} . \tag{2.18}$$

Therefore, in the presence of a magnetic monopole ($b \neq 0$) this condition excludes from the permissible range of values of $m$ the following set of odd integers:

$$m \notin \left\{-2(|b|+a) - 1 < m < 2(|b|-a) + 1\right\} . \tag{2.19}$$

Additionally, it is elementary to show that the right hand side of Eq. (2.17c) is always positive for all real values of $a$ and $b$ and for all odd integers $m$ and non-negative integers $n$. Thus, we can choose to write $\varepsilon_\theta^2 = \rho(\rho+1) + \frac{1}{4}$, where $\rho$ is a real physical parameter. Alternatively, $\varepsilon_\theta = \pm(\rho + \frac{1}{2})$. Therefore, for a given $\rho$ the odd integer $m$ could, in principle, assume any value in the range $m = \pm 1, \pm 3, \pm 5, ..., \pm \hat{m}$, where $\hat{m}$ is obtained from Eq. (2.17c) with $n = 0$ as the maximum positive odd integer satisfying

$$\hat{m} \leq -1 - 2|a| + 2\sqrt{\left(\rho + \frac{1}{2}\right)^2 + b^2} \ . \tag{2.20}$$

Merging this result with that in (2.18), we conclude that the admissible range of values of the odd integer $m$ when $b = 0$ is $-\hat{m} \leq m \leq \hat{m}$. However, for $b \neq 0$ we should exclude from this range the set of odd integers in (2.19). Now, for any odd integer $m$ in the permissible range obtained above, the non-negative integer $n$ is determined from Eq. (2.17c) as

$$n = \sqrt{\left(\rho + \frac{1}{2}\right)^2 + b^2} - \left|a + \frac{m}{2}\right| - \frac{1}{2} = 0, 1, 2, ... \tag{2.21}$$

Putting all of the above together, we obtain the following angular components of the spinor wavefunction

$$\Theta_s^\pm(\theta) = A_n \sqrt{\sin\theta} \, (1-x)^{\frac{1}{2}\left|a-b+\frac{m\mp 1}{2}\right|} (1+x)^{\frac{1}{2}\left|a+b+\frac{m\pm 1}{2}\right|} P_n^{(|a-b+\frac{m\mp 1}{2}|, |a+b+\frac{m\pm 1}{2}|)}(x), \tag{2.22}$$

where $m = \pm 1, \pm 3, \pm 5, ..., \pm \hat{m}$, excluding for $b \neq 0$ the set (2.19), and for each $m$ the non-negative integer $n$ is given by Eq. (2.21). Using the orthogonality relation of the Jacobi polynomials [8], one can easily verify that $\Theta_s^\pm$ obtained above makes the angular component in $f_\pm(\vec{r})$ orthonormal.

In the following section we complete the construction of the total solution space by obtaining the relativistic energy spectrum and the radial components of the spinor wave function that solve Eq. (2.10c) in the case where $V = 0$ and $W(r) \sim r$.

### III. SOLUTION OF THE RADIAL EQUATION

Transforming to the four component radial spinor $\begin{pmatrix} \tilde{R}_+ \\ \tilde{R}_- \end{pmatrix} = \begin{pmatrix} R_+ \\ \sigma_3 R_- \end{pmatrix}$ results in a mapping of the radial Dirac equation (2.10c) with $V = 0$ into the following

$$\begin{pmatrix} 1-\varepsilon & \lambda\left(-\frac{d}{dr}+\frac{\varepsilon_\theta}{r}+W\right) \\ \lambda\left(\frac{d}{dr}+\frac{\varepsilon_\theta}{r}+W\right) & -1-\varepsilon \end{pmatrix} \begin{pmatrix} \tilde{R}_+ \\ \tilde{R}_- \end{pmatrix} = 0. \quad (3.1)$$

This equation gives one radial component in terms of the other as

$$\tilde{R}_\mp = \frac{\lambda}{\varepsilon \pm 1}\left(\pm\frac{d}{dr}+\frac{\varepsilon_\theta}{r}+W\right)\tilde{R}_\pm, \quad (3.2)$$

where $\varepsilon \neq \mp 1$, respectively. This equation is referred to as the "kinetic balance" relation. Now, since $\varepsilon = +1$ ($\varepsilon = -1$) is an element of the positive (negative) energy spectrum, then this relation with the top (bottom) sign is not valid for the negative (positive) energy solutions. Our choice is to concentrate on obtaining the positive energy solutions. The negative energy solutions, on the other hand, are obtained from these positive energy solutions simply by the map

$$\varepsilon \to -\varepsilon, \ \varepsilon_\theta \to -\varepsilon_\theta, \ W \to -W, \ \tilde{R}_\pm \leftrightarrow \tilde{R}_\mp. \quad (3.3)$$

One can easily verify that this map transforms the kinetic balance relation (3.2) with the top sign into the same but with the bottom sign. Now, to obtain the positive energy solutions we use Eq. (3.2) to eliminate the lower radial component in favor of the upper in Eq. (3.1) resulting in the following Schrödinger-like second order differential equation

$$\left[-\frac{d^2}{dr^2}+\frac{\varepsilon_\theta(\varepsilon_\theta+1)}{r^2}+W^2-\frac{dW}{dr}+2\varepsilon_\theta\frac{W}{r}-\frac{\varepsilon^2-1}{\lambda^2}\right]\tilde{R}_+(r)=0. \quad (3.4a)$$

This shows that $\varepsilon_\theta$ plays the role of the angular momentum quantum number. The corresponding negative energy equation is obtained by using the kinetic balance relation (3.2) to eliminate the upper radial component in Eq. (3.1) giving

$$\left[-\frac{d^2}{dr^2}+\frac{\varepsilon_\theta(\varepsilon_\theta-1)}{r^2}+W^2+\frac{dW}{dr}+2\varepsilon_\theta\frac{W}{r}-\frac{\varepsilon^2-1}{\lambda^2}\right]\tilde{R}_-(r)=0. \quad (3.4b)$$

This equation could have also been obtained by the action of the map (3.3) on Eq. (3.4a). Moreover, in the language of supersymmetric quantum mechanics [9], these two equations are associated with the superpartner potentials (superpotentials) $\mathcal{W}^2 \pm \mathcal{W}'$, where $\mathcal{W} = W + \frac{\varepsilon_\theta}{r}$. For the Dirac-Oscillator problem we take $W(r)$ to be linear in the radial coordinate as $W = \omega^2 r$, where the real parameter $\omega$ is the oscillator frequency. Consequently, Eq. (3.4a) becomes

$$\left[-\frac{d^2}{dr^2}+\frac{\varepsilon_\theta(\varepsilon_\theta+1)}{r^2}+\omega^4 r^2+\omega^2(2\varepsilon_\theta-1)-\frac{\varepsilon^2-1}{\hbar^2}\right]\tilde{R}_+(r)=0. \tag{3.4a}'$$

Comparing this equation with that of the well-known nonrelativistic three-dimensional isotropic oscillator,

$$\left[-\frac{d^2}{dr^2}+\frac{\ell(\ell+1)}{r^2}+\omega^4 r^2-2E\right]\Psi(r)=0, \tag{3.5}$$

gives, by correspondence, the following map between the parameters of the two problems

$$\ell \to \begin{cases}\varepsilon_\theta \\ -\varepsilon_\theta-1\end{cases},\; E \to \frac{\varepsilon^2-1}{2\hbar^2}-\omega^2\left(\varepsilon_\theta-\tfrac{1}{2}\right) \tag{3.6}$$

The top (bottom) choice of the $\ell$ map corresponds to positive (negative) values of $\varepsilon_\theta$. Using this parameter map in the well-known nonrelativistic energy spectrum, $E_{k\ell} = \omega^2(2k+\ell+3/2)$ [10], gives the following relativistic positive energy spectrum

$$\varepsilon^+_{knm}=\sqrt{1+4\hbar^2\omega^2(k+\varepsilon_\theta+1/2)}\;,\;\varepsilon_\theta>0 \tag{3.7a}$$

$$\varepsilon^-_k=\sqrt{1+4\hbar^2\omega^2 k}\;,\;\varepsilon_\theta<0 \tag{3.7b}$$

where $k=0,1,2,..$ and the dependence on the integers $n$ and $m$ in (3.7a) comes form $\varepsilon_\theta$ as given by Eq. (2.17c). The lowest positive energy state, where $\varepsilon=+1$, occurs for $\varepsilon_\theta<0$ and $k=0$. In the absence of the magnetic monopole ($b=0$), $\varepsilon_\theta$ has a simpler expression and the energy spectrum (3.7a) reduces to

$$\varepsilon^+_{knm}=\begin{cases}\sqrt{1+4\hbar^2\omega^2(k+n+3/2)} & ,-1-2a<m<1-2a \\ \sqrt{1+4\hbar^2\omega^2(k+n+\left|a+\tfrac{m}{2}\right|+1)} & ,\text{otherwise}\end{cases} \tag{3.7a}'$$

The upper radial component $\tilde{R}_+$ of the positive energy solution is obtained using the same parameter map (3.6) into the nonrelativistic wavefunction [10]

$$\Psi_k(r)=A^\ell_k(\omega r)^{\ell+1}e^{-\omega^2 r^2/2}L^{\ell+\frac{1}{2}}_k(\omega^2 r^2), \tag{3.8}$$

where $A^\ell_k=\sqrt{2\omega\Gamma(k+1)/\Gamma(k+\ell+\tfrac{3}{2})}$, giving

$$\tilde{R}_+=e^{-\omega^2 r^2/2}\begin{cases}A^+_k(\omega r)^{\varepsilon_\theta+1}L^{\varepsilon_\theta+\frac{1}{2}}_k(\omega^2 r^2) & ,\varepsilon_\theta>0 \\ A^-_k(\omega r)^{-\varepsilon_\theta}L^{-\varepsilon_\theta-\frac{1}{2}}_k(\omega^2 r^2) & ,\varepsilon_\theta<0\end{cases} \tag{3.9}$$

where $A^\pm_k=\sqrt{2\omega\Gamma(k+1)/\Gamma(k\pm\varepsilon_\theta\pm\tfrac{1}{2}+1)}$. Using the kinetic balance relation (3.2) and the differential and recursion properties of the Laguerre polynomials [8], we obtain the (positive energy) lower radial spinor component as follows

$$\tilde{R}_-=2\hbar\omega e^{-\omega^2 r^2/2}\begin{cases}\dfrac{k+\varepsilon_\theta+1/2}{\varepsilon^+_{knm}+1}A^+_k(\omega r)^{\varepsilon_\theta}L^{\varepsilon_\theta-\frac{1}{2}}_k(\omega^2 r^2) & ,\varepsilon_\theta>0 \\ -\dfrac{1}{\varepsilon^-_k+1}A^-_k(\omega r)^{-\varepsilon_\theta+1}L^{-\varepsilon_\theta+\frac{1}{2}}_{k-1}(\omega^2 r^2) & ,\varepsilon_\theta<0\end{cases} \tag{3.10}$$

and $\tilde{R}_-=0$ for $k=0$ and $\varepsilon_\theta<0$, which corresponds to the lower bound of the spectrum where $\varepsilon=+1$. This is also supported by supersymmetric quantum mechanics where the superpotentials $\mathcal{W}^2\pm\mathcal{W}'$ in Eqs. (3.4a,b) are isospectral (i.e., share the same eigenvalue $\frac{\varepsilon^2-1}{\hbar^2}$) except for the zero eigenvalue ground state [9] which belongs only to the super-potential $\mathcal{W}^2 - \mathcal{W}'$ of Eq. (3.4a) for $\tilde{R}_+$. The radial spinor wavefunction associated with the lower bound of the positive energy spectrum, where $\varepsilon = +1$, $k = 0$ and $\varepsilon_\theta < 0$, is

$$\begin{pmatrix} \tilde{R}_+ \\ \tilde{R}_- \end{pmatrix} = \sqrt{2\omega/\Gamma(-\varepsilon_\theta + \tfrac{1}{2})}(\omega r)^{-\varepsilon_\theta} e^{-\omega^2 r^2/2} \begin{pmatrix} 1 \\ 0 \end{pmatrix}. \tag{3.11}$$

Finally, we obtain the total positive-energy four-component spinor wavefunction as follows:

$$\psi(\vec{r}) = \frac{1}{r\sqrt{\sin\theta}} \begin{pmatrix} if_+ \\ f_- \end{pmatrix} = \frac{1}{r\sqrt{\sin\theta}} \begin{pmatrix} i\Lambda g^+ \\ \Lambda g^- \end{pmatrix} = \frac{e^{\tfrac{i}{2}m\phi}}{r\sqrt{2\pi\sin\theta}} \begin{pmatrix} i\Lambda\begin{pmatrix} \Theta^+ \\ \Theta^- \end{pmatrix}\tilde{R}_+ \\ \Lambda\begin{pmatrix} \Theta^+ \\ -\Theta^- \end{pmatrix}\tilde{R}_- \end{pmatrix}, \tag{3.12}$$

In the following section we verify that the well-known spherically symmetric result (the pure Dirac-Oscillator) is a special case of our findings. Moreover, we obtain the non-relativistic limit and show that it agrees with nonrelativistic results reported elsewhere.

### IV. SPHERICAL SYMMETRY AND THE NONRELATIVISTIC LIMIT

It is straight-forward to verify that in the spherically symmetric case ($a = b = 0$) Equation (2.17c) gives

$$\varepsilon_\theta = \pm\left(n + \frac{|m|+1}{2}\right) = \pm 1, \pm 2, \pm 3, \ldots \tag{4.1}$$

Thus, $\rho = n + \frac{|m|}{2} = \frac{1}{2}, \frac{3}{2}, \frac{5}{2}, \ldots$ and $\varepsilon_\theta$ becomes the spin-orbit quantum number which is customarily referred to by the symbol $\kappa$

$$\varepsilon_\theta = \kappa = \pm\left(j + \tfrac{1}{2}\right) = \begin{cases} +j+\tfrac{1}{2}, & \ell = j+\tfrac{1}{2} \\ -j-\tfrac{1}{2}, & \ell = j-\tfrac{1}{2} \end{cases}, \tag{4.2}$$

where $\ell$ is the orbital angular momentum quantum number and $j$ is the total angular momentum (orbital plus spin), which is equal to $\rho$. Moreover, we can write $n = \rho - \frac{|m|}{2} = j - \frac{|m|}{2} = 0, 1, 2, \ldots$ Thus,

$$\tfrac{m}{2} = -j, -j+1, \ldots, j-1, j. \tag{4.3}$$

This range of values of $m$ is also obtainable using Eq. (2.20). By substituting these results in (3.7) we recover the familiar positive energy relativistic spectrum for the pure Dirac-Oscillator problem [2]

$$\varepsilon^+_{kj} = \sqrt{1 + 4\hbar^2\omega^2(k+j+1)}, \quad \varepsilon_\theta > 0 \tag{4.4a}$$

$$\varepsilon^-_k = \sqrt{1 + 4\hbar^2\omega^2 k}, \quad \varepsilon_\theta < 0 \tag{4.4b}$$

The total angular component of the spinor wavefunction could now be written as

$$Y^\pm_s(\theta,\phi) = \frac{\sqrt{\Gamma(j+\tfrac{1}{2}|m|+1)\Gamma(j-\tfrac{1}{2}|m|+1)}}{\sqrt{2^{|m|+1}\pi}\,\Gamma(j+\tfrac{1}{2})} \sqrt{\sin\theta}\,(1-x)^{\tfrac{|m\mp 1|}{4}}(1+x)^{\tfrac{|m\pm 1|}{4}} P^{(\tfrac{|m\mp 1|}{2},\tfrac{|m\pm 1|}{2})}_{j-\tfrac{1}{2}|m|}(x)\,e^{\tfrac{i}{2}m\phi}. \tag{4.5}$$

The radial components are obtained from Eqs. (3.9-3.11) with $\varepsilon_\theta$ as given by Eq. (4.2) above.

Finally, taking the nonrelativistic limit ($\hbar \to 0$) of the energy spectrum in (3.7a) and noting that in this limit $\varepsilon \to 1 + \hbar^2 E$, where $E$ is the nonrelativistic energy, the non-relativistic spectrum is obtained as

$$E_{knm} = 2\omega^2 \left[ k + \tfrac{1}{2} + \sqrt{\left(n + \tfrac{1}{2}\left|a - b + \tfrac{m\mp 1}{2}\right| + \tfrac{1}{2}\left|a + b + \tfrac{m\pm 1}{2}\right| + \tfrac{1}{2}\right)^2 - b^2} \right] \qquad (4.6)$$

This agrees with the findings in Appendix C of [11] for a charged particle in the presence of Aharonov-Bohm and magnetic monopole potentials but with the Coulomb interaction replaced by the oscillator. When comparing (4.6) with the results in [11] one should note that the azimuth phase quantum number there should be identified not with the odd integer $m$ above but with $\frac{m\pm 1}{2} = 0, \pm 1, \pm 2, \ldots$.